\documentclass[12pt]{iopart}
\usepackage{feynmp-auto}
\usepackage{iopams}
\usepackage{graphicx}
\usepackage{float}
\usepackage{subfig}
\begin{document}

\title[Green's function approach to the Bose-Hubbard model with disorder]{Green's function approach to the Bose-Hubbard model with disorder}

\author{R S Souza$^{1,*}$, Axel Pelster$^{2,\dag}$, and F E A dos Santos$^{1,\ddag}$}

\address{$^1$Departamento de Física, Universidade Federal de São Carlos, 13565-905 São Carlos, SP, Brazil}
\address{$^2$Physics Department and Research Center OPTIMAS, Technische Universität Kaiserslautern, 67663 Kaiserslautern, Germany}
\eads{\mailto{$^*$renan@df.ufscar.br}, \mailto{$^\dag$axel.pelster@physik.uni-kl.de}, and \mailto{$^\ddag$santos@ufscar.br}}
\vspace{10pt}
\begin{indented}
\item[]\today
\end{indented}

\begin{abstract}
We analyse the distinction between the three different ground states presented by a system of spinless bosons with short-range interactions submitted to a random potential using the disordered Bose-Hubbard model. The criteria for identifying the superfluid, the Mott-insulator, and the Bose-glass phases at finite temperatures are discussed for small values of the kinetic energy associated with the tunnelling of particles between potential wells. Field theoretical considerations are applied in order to construct a diagrammatic hopping expansion to the finite-temperature Green's function. By performing a summation of subsets of diagrams we are able to find the condition to the long-range correlations which leads to the phase boundary between superfluid and insulating phases. The perturbative expression to the local correlations allows us to calculate an approximation to the single-particle density of states of low-energy excitations in the presence of small hopping, which characterizes unambiguously the distinction between the Mott-insulator and the Bose-glass phases. We obtain the phase diagram for bounded on-site disorder. It is demonstrated that our analysis is capable of going beyond the mean-field theory results for the classification of these different ground states.
\end{abstract}

\section{Introduction}\label{S1}

Since the seminal paper of M. P. Fisher \textit{et al.} \cite{fisher1989boson}, the study of interacting bosonic particles in random potentials has become an active field of research. A key aspect of this system is the interplay between localization and superfluidity produced  by the combined effect of randomness and interaction. The experimental realization of Bose-Einstein condensation  in ultracold atomic gases together with the precise control presented by optical lattice experiments provided a unique possibility to investigate fundamental questions about this problem \cite{bloch2005ultracold,lye2005bose,fallani2007ultracold}. The random potential can be experimentally achieved, for instance, with magnetic wire traps \cite{wang2004disordered,schumm2005atom}, where imperfections of the wire produce local disorder, or even by using speckle laser fields, where a diffuse laser front creates the random lattice \cite{clement2005suppression,billy2008direct}. Perhaps the most pronounced phenomenon exhibited by this system is the superfluid to insulator quantum phase transition. 

The theoretical description of spinless bosons with short-range interactions moving in random external potentials is usually based on the disordered Bose-Hubbard Hamiltonian (BHH)
\begin{equation}\label{eq1}
\hat{H}_{BH}=\frac{U}{2}\sum_i\hat{n}_i(\hat{n}_i-1)-\sum_i (\mu-\epsilon_i)\hat{n}_i-J\sum_{\langle ij \rangle}\hat{a}^\dagger_i\hat{a}_j,
\end{equation} 
where $\hat{a}^\dagger_i$ and $\hat{a}_i$ are the bosonic creation and annihilation operators fulfilling the canonical commutation relations, $\hat{n}_i=\hat{a}^\dagger_i\hat{a}_i$ denotes the number operator, $\mu$ is the chemical potential, and the sum $\langle ij \rangle$ runs over nearest neighbours. In addition, the interaction between two particles at the same lattice site is parametrized by the energy $U$, the hopping parameter $J$ corresponds to the kinetic energy associated with the tunnelling of a particle from a lattice site to one of its first neighbours, and the on-site energies $\epsilon_i$ represent local imperfections which we assume to be uncorrelated at different sites and thus randomly spread over the lattice obeying some probability distribution $p(\{\epsilon_i\})$. A detailed analysis \cite{zhou2010construction} demonstrates that the disorder generated,, for instance, by laser speckles would make all parameters random and depending on the speckle potential distribution. However, we will assume that \eref{eq1} is valid as current experiments show that by choosing an appropriate holographic mask when imaging under a quantum gas microscope one can create arbitrary potential landscapes \cite{bakr2009quantum}. Additionally, it was shown in \cite{bender2018customizing} that the speckle intensity can be customised to generate a wide variate of speckle patterns including a uniform distributions. 

The competition between the system parameters gives rise to different phase transitions. If the hopping energy is much larger when compared to the interaction energy, the atoms can move without viscosity over the system's volume and the ground state is superfluid. In the opposite case, in which interactions dominate over the tunnelling energy, a finite number of atoms becomes localized around each potential minimum configuring a Mott-insulator state. The superfluid to Mott insulator transition was directly tested by observing the multiple matter wave interference pattern presented in absorption pictures of time-of-flight measurements taken for different lattice potential depths \cite{greiner2002quantum}. When disorder is introduced, a third phase intervenes between the latter two: the Bose-glass phase. This phase consists of rare superfluid regions inside an insulating background and for sufficiently strong disorder strength it can even destroy the Mott-insulator state. The superfluid-Bose glass transition has dynamically been probed using a quantum quench of disorder in an ultracold gas at non-zero temperature and measuring its excitations in an experiment by Meldgin \textit{et al.} \cite{meldgin2016probing}. Although some attributes of the Bose-glass phase are well understood, detailed information concerning its phase boundary as well as on the nature of its low-lying excitations are still lacking both from experimental and theoretical points of view.

The analytic form of the BHH eigenstates and eigenenergies cannot be
directly computed in general. Thus, with the aim of reaching predictions for the phase boundaries of the transitions, numerical methods such as Monte-Carlo simulations \cite{astrakharchik2002superfluidity,capogrosso2007phase,meier2012quantum,zhang2015equilibrium,ng2015quantum,de2018properties} and stochastic \cite{bissbort2009stochastic,bissbort2010stochastic} as well as local \cite{thomson2016measuring} mean-field techniques have been applied, while analytic investigations have mostly been confined to the method of mean-field theory \cite{krutitsky2006mean,buonsante2007mean,pisarski2011application}. More recently, the application of field-theoretical approaches proved to be efficient in the analysis of the above described quantum phase transitions providing precise results when compared to Monte-Carlo simulations in the pure case \cite{dos2009quantum,bradlyn2009effective,dos2011ginzburg,ohliger2013green}. The critical exponents for the superfluid to Bose-glass phase transition have been numerically calculated in \cite{yao2014critical}. Analytical information concerning the impact of temperature on the phase boundary between Mott insulator and Bose glass was obtained in 
\cite{krutitsky2006mean}. In this study, the zero temperature characteristics of these two states were tested in a finite-temperature theory by analysing the single-particle density of states. Unlike the superfluid state, both insulating phases are distinguished by the absence of off-diagonal long-range order. However, the existence of a finite energy gap for particle-hole excitations in the Mott-insulator phase leads to a vanishing density of states at zero energy, in contrast to the Bose-glass phase which presents a gapless single-particle excitation spectrum and consequently a finite zero-energy density of states \cite{fisher1989boson}. Even though such distinctions were tested on a finite-temperature theory, the influence of finite values of the tunnelling energy to their phase boundary remains to be studied. Therefore, it is important to investigate to which extent these definitions or similar ones still hold at finite temperatures and at least for small values of the hopping parameter. In this paper, we demonstrate that such an analysis can be performed by calculating corrections to the Green's function due to the hopping of particles. To this end, we develop a perturbative treatment to the BHH considering bounded on-site disorder. Using field theoretical considerations, similarly to \cite{dos2009quantum,bradlyn2009effective,dos2011ginzburg,ohliger2013green}, we construct a hopping parameter expansion to the two-point correlation function at finite temperatures. The phase boundary to the superfluid phase is identified by observing the divergence of the resummed expression of the correlation function,  while the phase boundary distinguishing the
Mott-insulator from the Bose-glass phase is computed by analysing the imaginary part of the same-site correlation function in real frequency space, which corresponds to the single-particle density of states.

In what follows, we first construct the perturbative expression to the Green's function in \sref{S2} and then proceed the investigation to obtain the superfluid to insulator phase boundary in \sref{S3}. In \sref{S4}, we analyse the imaginary part of the local Green's function in order to calculate the first relevant correction to the single-particle density of states. In \sref{S5}, we analyse the distribution of such a quantity to determine the Mott insulator to Bose glass phase boundary. A comparison of our results for the first Mott lobe with the numerical predictions of \cite{bissbort2009stochastic,bissbort2010stochastic,thomson2016measuring}  for the $2D$ and $3D$ cases at both zero and finite temperatures is presented in \sref{S6}. In \sref{S7}, we then make a summary of our findings and conclude that this method is capable of going beyond mean-field theory for the analysis of the phase transition.

\section{Perturbation theory}\label{S2}
We base our analysis on the single-particle Green's function as it characterizes the microscopic properties of the system. This function can be defined as the thermal average of the time-ordered product of the bosonic creation and annihilation operators in the Heisenberg representation
\begin{equation}
G_{ij}(\tau;\tau^\prime)=\langle \hat{\mathcal{T}}[\hat{a}_{i}(\tau)\hat{a}^\dagger_{j}(\tau^\prime)]\rangle,
\end{equation}
where $\hat{\mathcal{T}}$ is the time-ordering operator and the Heisenberg representation for an arbitrary Schrödinger operator $\hat{O}_s$ is defined as $\hat{O}(\tau)=\rme^{\tau\hat{H}}\hat{O}_s\rme^{-\tau\hat{H}}$, with $\hbar=1$. As one of our main interests is to describe the system at finite temperatures, we have used the Wick rotation $t\rightarrow -\rmi\tau$ to establish the imaginary-time formalism \cite{abrikosov1963methods,zinn1996quantum}, where $\tau$ is the so-called imaginary time.

As an exact diagonalization of the BHH is not possible, the Green's function as well as other important quantities of the system may be calculated perturbatively. To this end, we first consider the BHH as belonging to a general class of Hamiltonians composed of a local term plus a hopping term
\begin{equation}
\hat{H}=\sum_i \hat{H}_{0_i}-\sum_{ij}J_{ij}\hat{a}^\dagger_i\hat{a}_j,
\end{equation}
where $J_{ij}$ is symmetric in $i$ and $j$ and $J_{ii}=0$. The BHH is recovered by setting $\hat{H}_{0_i}=U\hat{n}_i(\hat{n_i}-1)/2-\mu_i\hat{n}_i$, where $\mu_i=\mu-\epsilon_i$, and considering hopping only between first neighbouring sites. Following field-theoretic considerations \cite{zinn1996quantum,kleinert2001critical,kleinert2009path}, we then include a source term to the Hamiltonian with the intention of explicitly breaking any global symmetries
\begin{equation}
\hat{H}(\tau)=\hat{H}-\sum_i\Big[\mathrm{j}_i(\tau)\hat{a}^\dagger_i+\mathrm{j}^*_i(\tau)\hat{a}_i\Big].
\end{equation}
Using the Dirac interaction picture, the initial value problem for the imaginary-time evolution operator takes the form
\begin{equation}
\frac{\partial\hat{\mathcal{U}}_I(\tau,\tau_0)}{\partial\tau}=-\hat{H}_I(\tau)\hat{\mathcal{U}}_I(\tau,\tau_0),\qquad\mbox{with}\qquad\hat{\mathcal{U}}_I(\tau_0,\tau_0)=1,
\label{eq5}
\end{equation}
where $\hat{H}_I(\tau)$ is the interaction picture representation of the hopping term plus the source term
\begin{equation}
\hat{H}_I(\tau)=-\sum_{ij}J_{ij}\hat{a}^\dagger_i(\tau)\hat{a}_j(\tau)-\sum_i\Big[\mathrm{j}_i(\tau)\hat{a}^\dagger_i(\tau)+\mathrm{j}^*_i(\tau)\hat{a}_i(\tau)\Big].
\end{equation}
\Eref{eq5} has a solution which is given by the Dyson series
\begin{equation}
\hat{\mathcal{U}}_I[\mathrm{j},\mathrm{j^*}](\tau,\tau_0)=\hat{\mathcal{T}}\exp\Bigg( -\int_{\tau_0} ^\tau \rmd\tau^\prime\hat{H}_I(\tau^\prime)\Bigg).
\end{equation}

Starting from the fully localized case, $J_{ij}=0$, the hopping-free partition function can be written as 
\begin{equation}
\mathcal{Z}_0[\mathrm{j},\mathrm{j^*}]=\tr\Big(\rme^{-\beta\hat{H}_0}\hat{\mathcal{U}}_s[\mathrm{j},\mathrm{j^*}](\beta,0)\Big),
\end{equation}
where
\begin{equation}
\hat{\mathcal{U}}_s[\mathrm{j},\mathrm{j^*}](\beta,0)=\hat{\mathcal{T}}\exp\Bigg( \int_{0} ^\beta\rmd\tau\sum_i\Big[\mathrm{j}_i(\tau)\hat{a}^\dagger_i(\tau)+\mathrm{j}^*_i(\tau)\hat{a}_i(\tau)\Big]\Bigg),
\end{equation}
with $\beta=1/k_BT$, where $k_B$ is the Boltzmann constant and $T$ the temperature. Using the semi-group property of the imaginary-time evolution operator \cite{dos2011ginzburg}, we can express the full partition function as a power series in the hopping matrix elements given by
\begin{equation}
\mathcal{Z}[\mathrm{j},\mathrm{j^*}]=\exp\Bigg(\sum_{ij}J_{ij}\int_{0} ^\beta\rmd\tau\frac{\delta^2}{\delta\mathrm{j}^*_i(\tau)\delta\mathrm{j}_j(\tau)}\Bigg)\mathcal{Z}_0[\mathrm{j},\mathrm{j^*}].
\label{eq10}
\end{equation}
With the same property it is possible to show that the Green's function can be calculated by taking functional derivatives of the full partition function with respect to the sources and then considering the limit where they vanish 
\begin{equation}
G_{ij}(\tau;\tau^\prime)=\frac{1}{\mathcal{Z}[\mathrm{j},\mathrm{j^*}]}\frac{\delta^{2}\mathcal{Z}[\mathrm{j},\mathrm{j^*}]}{\delta\mathrm{j}^*_{i}(\tau)\delta\mathrm{j}_{j}(\tau^\prime)}\Big\vert_{\mathrm{j}=\mathrm{j}^*=0}.
\label{eq11}
\end{equation}

Thus, the problem is reduced to finding the expression of $\mathcal{Z}_0$ and then taking its functional derivatives with respect to the sources in order to account for the hopping contributions. This calculation is simplified if we notice that in the local case the hopping-free partition function becomes a product of single-site contributions
\begin{equation}
\mathcal{Z}_0[\mathrm{j},\mathrm{j^*}]=\prod_i \mathcal{Z}_{0i}[\mathrm{j},\mathrm{j^*}].
\end{equation}
As a consequence, the free energy $W_0=-\beta F_0=\ln(\mathcal{Z}_0)$ also becomes local. Hence, in the zero hopping case we have
\begin{equation}
W_0[\mathrm{j},\mathrm{j^*}]=\sum_i W_{0i}[\mathrm{j}_i,\mathrm{j}^*_i],
\label{eq13}
\end{equation}
and we can expand each local term in the summation as a series in the sources
\begin{equation}
W_{0i}[\mathrm{j}_i,\mathrm{j}_i^*]=W^{(0)}_{0i}+\int_{0} ^\beta\rmd\tau\int_{0} ^\beta\rmd\tau^\prime \mathrm{j}^*_i(\tau)W^{(2)}_{0i}(\tau;\tau^\prime)\mathrm{j}_i(\tau^\prime)
+\cdots,
\label{eq14}
\end{equation}
where the functions $W^{(2n)}_{0i}(\tau_1,\ldots,\tau_n;\tau^\prime_1,\ldots,\tau^\prime_n)$ are the so-called local $2n$-point correlation functions. We focus our analysis on the calculation of the $2$-point correlations. For this purpose, it is convenient to use a diagrammatic notation similarly to \cite{metzner1991linked}. To construct such a notation, we associate each $2n$-point function $W^{(2n)}_{0i}$ with a vertex labelled with a site index $i$ and containing $n$ entering lines and $n$ exiting lines which correspond to the imaginary-time variables $\tau^\prime_n$ and $\tau_n$, respectively. Thus, the terms shown in the above equation are respectively represented by
\begin{equation}
\begin{fmffile}{f0}
W^{(0)}_{0i}=\parbox{2mm}{\begin{fmfgraph*}(4,4)
		\fmfleft{i}
			\fmfright{o}
			\fmf{phantom}{i,v,o}
			\fmfv{l=$i$,l.a=-90}{v}
			\fmfdot{v}
		\end{fmfgraph*}}\qquad\mbox{and}\qquad
W^{(2)}_{0i}(\tau;\tau^\prime)=\quad\quad
		\parbox{20mm}{\begin{fmfgraph*}(55,2)
			\fmfleft{i}
			\fmfright{o}
			\fmf{fermion}{i,v,o}
			\fmfv{l=$\tau^\prime$}{i}
			\fmfv{l=$i$,l.a=-90}{v}
			\fmfv{l=$\tau$}{o}
			\fmfdot{v}
			\end{fmfgraph*}}\quad.
\end{fmffile}
\end{equation}
The notation can be shortened by suppressing both the vertex label when summing over all lattice sites, and each inward (outward) line label when multiplying by $\mathrm{j}_i$ ($\mathrm{j}_i^*$) and integrating from $0$ to $\beta$ in the imaginary-time variable. Thereafter, we can rewrite \eref{eq13} together with \eref{eq14} as a sum of 1-vertex diagrams
\begin{equation}\label{eq16}
\begin{fmffile}{f2}
W_0[\mathrm{j},\mathrm{j}^*]=
		\parbox{2mm}{\begin{fmfgraph}(4,4)
		\fmfiv{d.sh=circle,d.f=1,d.si=4}{c}
		\end{fmfgraph}}
		+\parbox{20mm}{\begin{fmfgraph}(55,2)
			\fmfleft{i}
			\fmfright{o}
			\fmf{fermion}{i,v,o}
			\fmfdot{v}
			\end{fmfgraph}}
			+\cdots.
\end{fmffile} 
\end{equation}
As our approach is applied only to the 2-point diagram, we choose not to show the subsequent terms in this expansion, which would consist of all the diagrams with an even number greater than $2$ of external lines joined by a single vertex. The next diagram in the summation, the $4$-point diagram, becomes important, for instance, in the effective-action approach of \cite{dos2009quantum,bradlyn2009effective,dos2011ginzburg,ohliger2013green}.

To compute the hopping corrections to the free energy, we must apply \eref{eq10} to $\mathcal{Z}_0=\exp(W_0)$ and then take the logarithm of the result. In diagrammatic notation, the hopping matrix can be denoted by an internal line between two vertices,
\begin{equation}
\begin{fmffile}{f3}
J_{ij}=\quad\quad
		\parbox{20mm}{\begin{fmfgraph*}(55,2)
			\fmfleft{i}
			\fmfright{o}
			\fmf{fermion}{i,o}
			\fmfv{l=$i$}{i}
			\fmfv{l=$j$}{o}
			\end{fmfgraph*}}\quad,
\end{fmffile}
\end{equation}
and the effect of the functional derivatives with respect to the sources $\mathrm{j}_i$ and $\mathrm{j}^*_i$ on a local diagram is the introduction of an index $i$ to its central vertex and the addition of an imaginary time variable $\tau$ to its inward or outward lines, respectively. The result of the full operator acting on products of diagrams is to generate different diagrams by joining an inward open line of one diagram with the outward open line of another. The linked-cluster theorem \cite{irving1984methods,gelfand1990perturbation} assures that only the connected diagrams will contribute to $W$. Therefore, the hopping expansion to the free energy is given by
\begin{equation}
\fl
\begin{fmffile}{f4}
W[\mathrm{j},\mathrm{j}^*]
		=
		\parbox{2mm}{\begin{fmfgraph}(4,4)
		\fmfiv{d.sh=circle,d.f=1,d.si=4}{c}
		\end{fmfgraph}}
		+
		\parbox{20mm}{\begin{fmfgraph*}(55,2)
			\fmfleft{i}
			\fmfright{o}
			\fmf{fermion}{i,v,o}
			\fmfdot{v}
			\end{fmfgraph*}}
			+
			\parbox{30mm}{\begin{fmfgraph*}(85,2)
			\fmfleft{i}
			\fmfright{o}
			\fmf{fermion}{i,v1,v2,o}
			\fmfdot{v1,v2}
			\end{fmfgraph*}}
			+
			\parbox{40mm}{\begin{fmfgraph*}(110,2)
			\fmfleft{i}
			\fmfright{o}
			\fmf{fermion}{i,v1,v2,v3,o}
			\fmfdot{v1,v2,v3}
			\end{fmfgraph*}}
			+\cdots,
\label{eq18}
\end{fmffile}
\end{equation}
where we have considered only tree-level corrections to the $2$-point diagrams, which are the ones that will be essential to our analysis. Note that \eref{eq18} differs from \eref{eq16} as it contains all simple-chain diagrams with two external lines, where the order of each diagram in the hopping approximation is determined by the number of internal lines between the vertices. By using \eref{eq18}, to write $\mathcal{Z}[\mathrm{j},\mathrm{j}^*]=\exp(W[\mathrm{j},\mathrm{j}^*])$, and then applying \eref{eq11} we obtain the Green's function
\begin{equation}
\fl
\begin{fmffile}{f5}
G_{ij}(\tau;\tau^\prime)=\delta_{ij}
		\parbox{20mm}{\begin{fmfgraph*}(55,25)
			\fmfleft{i}
			\fmfright{o}
			\fmf{fermion}{i,v,o}
			\fmfv{l=$\tau^\prime$,l.a=-90}{i}
			\fmfv{l=$i$,l.a=-90}{v}
			\fmfv{l=$\tau$,l.a=-90}{o}
			\fmfdot{v}
			\end{fmfgraph*}}
			+
			\parbox{30mm}{\begin{fmfgraph*}(85,25)
			\fmfleft{i}
			\fmfright{o}
			\fmf{fermion}{i,v1,v2,o}
			\fmfdot{v1,v2}
			\fmfv{l=$\tau^\prime$,l.a=-90}{i}
			\fmfv{l=$i$,l.a=-90}{v1}
			\fmfv{l=$j$,l.a=-90}{v2}
			\fmfv{l=$\tau$,l.a=-90}{o}
			\end{fmfgraph*}}
			+
			\parbox{40mm}{\begin{fmfgraph*}(110,25)
			\fmfleft{i}
			\fmfright{o}
			\fmf{fermion}{i,v1,v2,v3,o}
			\fmfdot{v1,v2,v3}
			\fmfv{l=$\tau^\prime$,l.a=-90}{i}
			\fmfv{l=$i$,l.a=-90}{v1}
			\fmfv{l=$j$,l.a=-90}{v3}
			\fmfv{l=$\tau$,l.a=-90}{o}
			\end{fmfgraph*}}
			+
			\cdots,
\label{eq19}
\end{fmffile}
\end{equation}
which consists of a sum of all simple chain diagrams. With this expression we can now proceed to determine the phase boundaries for the quantum phase transitions.

\section{Superfluid phase boundary}\label{S3}

The evaluation of the diagrams introduced above involves an integration in the imaginary time variables. This process can be carried out by considering the transformation to the Matsubara frequency space
\begin{equation}
\eqalign{g(\omega_l)=\int^\beta_0\rmd\tau g(\tau)\rme^{\rmi\omega_l\tau},\\ g(\tau)=\frac{1}{\beta}\sum_{\omega_l=-\infty}^\infty g(\omega_l)\rme^{-\rmi\omega_l\tau},}
\end{equation}
with the bosonic Matsubara frequencies $\omega_l=2\pi l/\beta$, where $l\in\mathbb{Z}$.
It turns out that working in the frequency space simplifies our calculations as the system that we are considering presents time-translation invariance. The diagrammatic expansion maintains the same form in Matsubara space.

As shown in \cite{kleinert2001critical}, the full correlation functions can be decomposed into connected correlation functions. In our theory, this decomposition for the $2$-point function gives
\begin{equation}
\fl
\eqalign{W_{0i}^{(2)}(\tau,\tau^\prime)=\frac{1}{Z_0(\mu_i)}\sum_{n=0}^\infty \rme^{-\beta f_n(\mu_i)}\Big(\Theta(\tau-\tau^\prime)&(n+1)\rme^{(\tau-\tau^\prime)[f_n(\mu_i)-f_{n+1}(\mu_i)]}\\
&+\Theta(\tau^\prime-\tau)n\rme^{(\tau^\prime-\tau)[f_n(\mu_i)-f_{n-1}(\mu_i)]}\Big),}
\end{equation}
where 
\begin{equation}
\eqalign{Z_0(\mu_i)&=\sum_{n=0}^\infty \rme^{-\beta f_n(\mu_i)},\\ f_n(\mu_i)&=\langle n|\hat{H}_{0_i}|n\rangle\\&=\case{U}{2}n(n-1)-\mu_in,}
\label{eq22}
\end{equation}
and $\Theta(\tau)$ is the Heaviside step function. In the Matsubara representation, we get
\begin{equation}
W_{0i}^{(2)}(\omega_{l_1},\omega_{l_2})=\beta\delta_{\omega_{l_1}\omega_{l_2}}g_i(\omega_{l_1}),
\label{eq23}
\end{equation}
where $\delta_{\omega\omega^\prime}$ is the Kronecker delta and we have defined
\begin{equation}
g_i(\omega_l)=\frac{1}{Z_0(\mu_i)}\sum_{n=0}^\infty \rme^{-\beta f_n(\mu_i)}\Bigg(\frac{n}{\rmi\omega_l+\mu_i-U(n-1)}-\frac{n+1}{\rmi\omega_l+\mu_i-Un}\Bigg).
\label{eq24}
\end{equation} 

In order to deal with the random chemical potential, we consider the disorder to be frozen in time. Also, the magnitude of the local chemical potential at different lattice sites can be regarded as spatially uncorrelated, thus varying within a range where each value appears with a specific probability. This is equivalent to assume a lattice spacing which is much larger than the disorder correlation length. Accordingly, the local disorder $\epsilon_i$ is assumed to be characterized by a probability distribution $p(\epsilon_i)$ of some kind. As a consequence, it becomes necessary to define a disorder ensemble average which has the following form
\begin{equation}
\overline{G}=\prod_i\int^\infty_{-\infty}\rmd\epsilon_iG(\epsilon_i)p(\epsilon_i).
\end{equation} 

The transition to the superfluid phase is characterized by diverging long-range correlations \cite{zinn1996quantum,kleinert2001critical}. However, any finite order approximation in the hopping expansion \eref{eq19} of the Green's function is a power series in $J$ and therefore analytic. Hence, we must perform a summation of the infinite subset of chain diagrams, in the same manner as in \cite{ohliger2013green}. This task becomes straightforward if we make the transformation of the disorder average of \eref{eq19} to quasi-momentum space
\begin{equation}
\overline{G}(\bi{k},\omega_{l_1};\bi{k}^\prime,\omega_{l_2})=\sum_{ij}\overline{G}_{ij}(\omega_{l_1},\omega_{l_2})\exp[-\rmi(\bi{k}\cdot\bi{r}_i-\bi{k}^\prime\cdot\bi{r}_j)]. 
\end{equation}
The above expression can be rewritten as
\begin{equation}
\overline{G}(\bi{k},\omega_{l_1};\bi{k}^\prime,\omega_{l_2})=\beta\Big(\frac{2\pi}{a}\Big)^D\delta_{\omega_{l_1}\omega_{l_2}}\delta(\bi{k}-\bi{k}^\prime)\overline{G}(\bi{k},\omega_{l_1}),
\end{equation}
with
\begin{equation}
\overline{G}(\bi{k},\omega_{l_1})=\sum_{n=0}^\infty\big[\overline{g}_i(\omega_{l_1})\big]^{n+1}J(\bi{k})^n,
\end{equation}
where $J(\bi{k})=2J\sum_{\alpha=1}^D\cos(k_\alpha a)$ is the $D$-dimensional lattice dispersion. This equation consists of a geometric series which can be directly evaluated to
\begin{equation}
\overline{G}(\bi{k},\omega_{l_1})=\Bigg[\frac{1}{\overline{g}_i(\omega_{l_1})}-J(\bi{k})\Bigg]^{-1}.
\end{equation}

Note that the  imaginary part of $\overline{g}_i(\omega_{l_1})$ in \eref{eq24} vanishes for $\omega_{l_1}=0$. Furthermore, as phase transitions are governed by long-wavelength fluctuations \cite{zinn1996quantum,kleinert2001critical}, we must also set $\bi{k}=\bi{0}$. In this situation we find that the Green's function diverges in $J$ when
\begin{equation}
\frac{J}{U}=\Bigg[\int^{\mu+\infty}_{\mu-\infty}\rmd \mu_i\frac{p(\mu_i-\mu)}{zZ_0(\mu_i)}\sum_{n=0}^\infty\frac{n+1}{\frac{\mu_i}{U}-n}(\rme^{-\beta f_{n+1}(\mu_i)}-\rme^{-\beta f_n(\mu_i)})\Bigg]^{-1},
\label{eq30}
\end{equation}
where $z=2D$ is the lattice coordination number. This result for the phase boundary is exactly equal to the one obtained in \cite{krutitsky2006mean} with mean-field theory. For the uniform disorder distribution 
\begin{equation}
p(\epsilon)=\frac{1}{\Delta}\Big[\Theta\Big(\epsilon+\frac{\Delta}{2}\Big)-\Theta\Big(\epsilon-\frac{\Delta}{2}\Big)\Big],
\label{eq31}
\end{equation}
the phase boundary becomes
\begin{equation}
\frac{J}{U}=\Bigg[\frac{1}{z\Delta}\int^{\mu+\frac{\Delta}{2}}_{\mu-\frac{\Delta}{2}}\rmd \mu_i\frac{1}{Z_0(\mu_i)}\sum_{n=0}^\infty\frac{n+1}{\frac{\mu_i}{U}-n}(\rme^{-\beta f_{n+1}(\mu_i)}-\rme^{-\beta f_n(\mu_i)})\Bigg]^{-1}.
\label{eq32}
\end{equation}
\begin{figure}[!]
\centering
\subfloat[]{\includegraphics[width=0.48\textwidth]{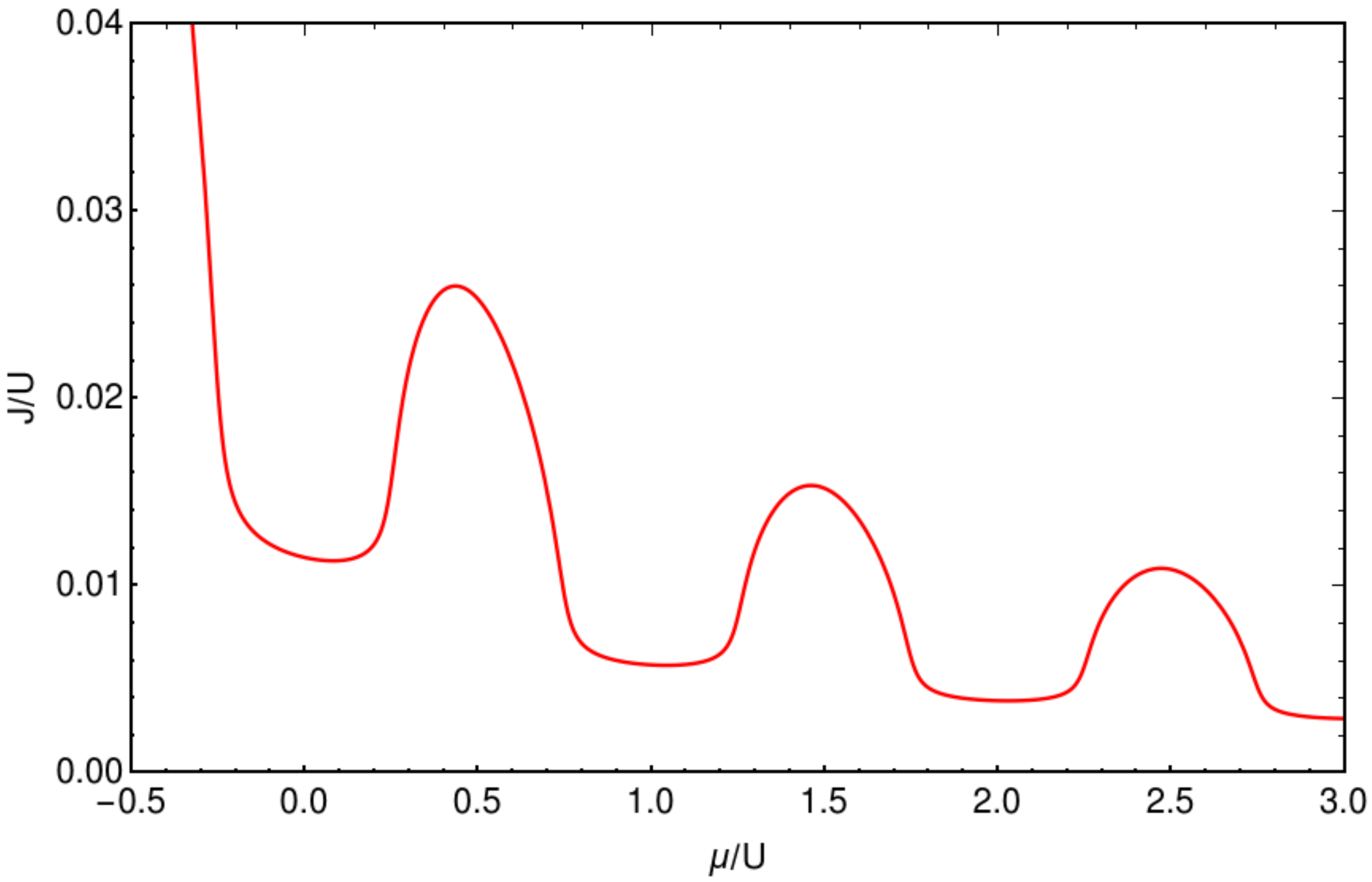}}
\hfill
\subfloat[]{\includegraphics[width=0.48\textwidth]{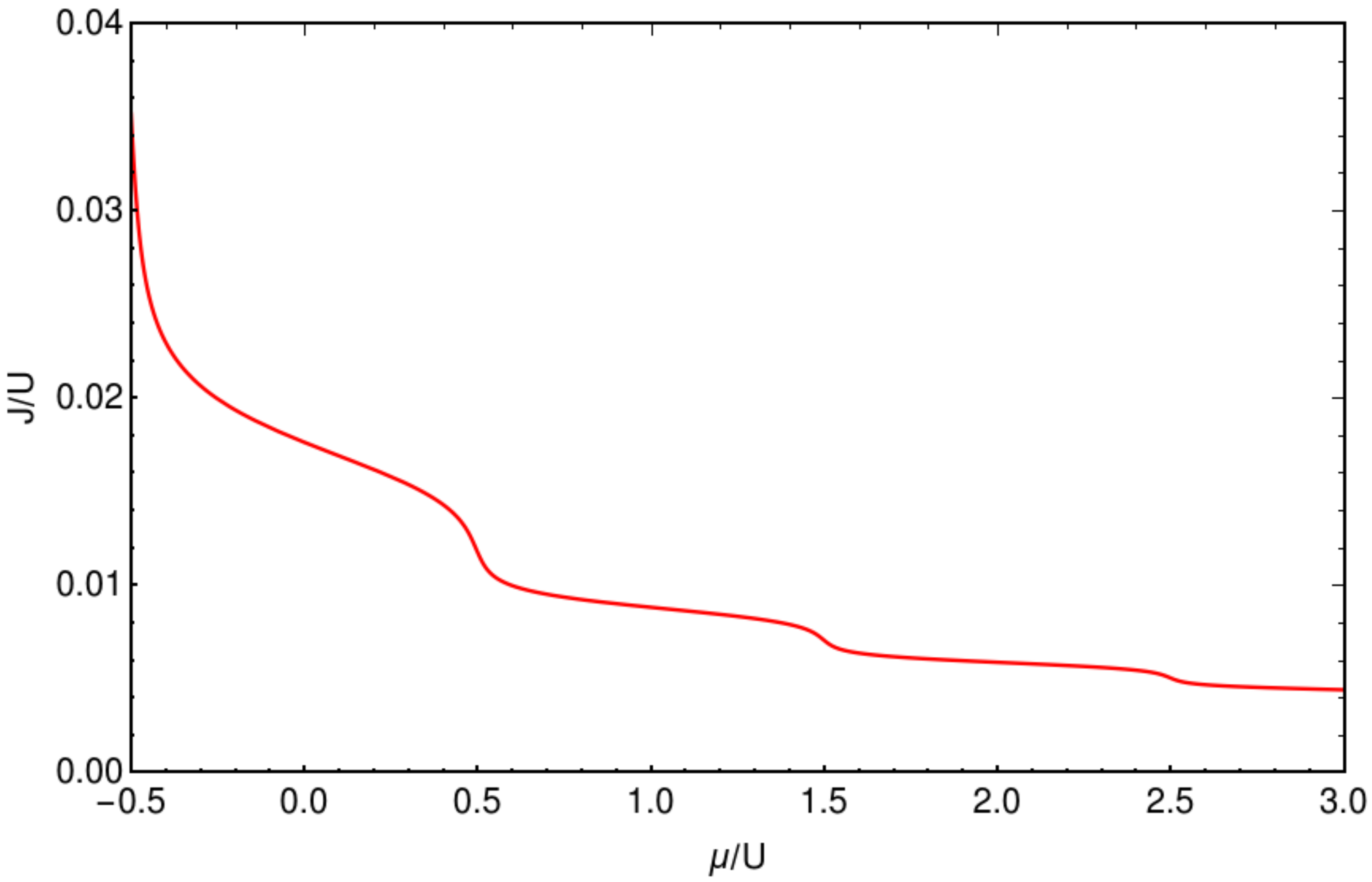}}
\caption{Phase boundary between superfluid and insulating phases obtained from \eref{eq32} for the finite temperature $k_BT/U=0.01$, with $z=6$, and for different values of the disorder parameter.  The disorder strength is $\Delta/U=0.5$ in (a) and $\Delta/U=1$ in (b).}\label{fig1}
\end{figure}

The result of the above equation for different disorder strengths can be observed in \fref{fig1}. There, we can see the phase boundary between superfluid and insulating phases at a finite temperature. Above the red line only the superfluid phase can exist. The region enclosed below this line corresponds to non-superfluid phases. We can see that, with an increase in the disorder strength, from \fref{fig1}(a) to \ref{fig1}(b), the phase boundary becomes smoother. However, no distinction between Mott insulator and Bose glass can be made in this diagram. As previously stated, such a distinction can be clarified by analysing the behaviour of the density of states for single excitations below the superfluid region, which is what we develop in the following.

\section{Single-particle density of states}\label{S4}

According to \cite{fisher1989boson}, for vanishing hopping, the density of states of the low-lying excitations in the well-localized regime of the Bose-glass phase is constant at zero excitation energy by virtue of the continuous distribution of the random potential. It was further stated that this situation should be sustained when the hopping parameter is made slightly positive resulting in a single-particle density of states also constant at zero energy. In light of these arguments, we now consider the hopping corrections of our perturbation theory to the local correlation function from which we can obtain the density of states of single excitations. 

In the same site case, $i=j$, the first-order term in \eref{eq19} vanishes due to the fact that there should be no possibility of a particle hopping from a site to itself, $J_{ii}=0$, and therefore the lowest relevant contribution to the local Green's function is of second order, resulting in the following diagrammatic expression
\begin{equation}\label{eq33}
\begin{fmffile}{f6}
G_{i}(\tau;\tau^\prime)=
		\parbox{22mm}{\begin{fmfgraph*}(55,2)
			\fmfleft{i}
			\fmfright{o}
			\fmf{fermion}{i,v,o}
			\fmfv{l=$\tau^\prime$,l.a=-90}{i}
			\fmfv{l=$i$,l.a=-90}{v}
			\fmfv{l=$\tau$,l.a=-90}{o}
			\fmfdot{v}
			\end{fmfgraph*}}
			+
			\parbox{42mm}{\begin{fmfgraph*}(110,20)
			\fmfleft{i}
			\fmfright{o}
			\fmf{fermion}{i,v1,v2,v3,o}
			\fmfdot{v1,v2,v3}
			\fmfv{l=$\tau^\prime$,l.a=-90}{i}
			\fmfv{l=$i$,l.a=-90}{v1}
			\fmfv{l=$i$,l.a=-90}{v3}
			\fmfv{l=$\tau$,l.a=-90}{o}
			\end{fmfgraph*}}
			+
			\cdots.
\end{fmffile}
\end{equation}
Note that by using local correlations we restrict our calculation to be valid only inside the non-superfluid part of the phase diagram. In Matsubara space, \eref{eq33} can be written as
\begin{equation}
\eqalign{G_{i}(\omega_{l_1},\omega_{l_2})&=\beta\delta_{\omega_{l_1},\omega_{l_2}}G_{i}(\omega_{l_1})\\ &=\beta\delta_{\omega_{l_1},\omega_{l_2}}\Big(g_i(\omega_{l_1})+\sum_jJ_{ij}J_{ji}[g_i(\omega_{l_1})]^2g_j(\omega_{l_1})\Big)+\cdots,}
\label{eq34}
\end{equation}
which explicitly reads
\begin{equation}
\fl
\eqalign{G_{i}(\omega_{l_1})=&\frac{1}{Z_0(\mu_i)}\sum_{n=0}^\infty \rme^{-\beta f_n(\mu_i)}\Bigg(\frac{n}{\rmi\omega_{l_1}+\mu_i-U(n-1)}-\frac{n+1}{\rmi\omega_{l_1}+\mu_i-Un}\Bigg)
\\
&+\sum_jJ_{ij}J_{ji}\sum_{n,m=0}^\infty \Bigg(\frac{\rme^{-\beta f_n(\mu_i)}}{Z_0(\mu_i)}\Bigg)^2\frac{\rme^{-\beta f_m(\mu_j)}}{Z_0(\mu_j)}
\\
&\quad\times\frac{\partial}{\partial\mu_i}\Bigg(
\frac{a^{(1)}_{n,m}(\mu_i,\mu_j)}{\rmi\omega_{l_1}+\mu_i-U(n-1)}+
\frac{a^{(2)}_{n,m}(\mu_i,\mu_j)}{\rmi\omega_{l_1}+\mu_i-Un}\Bigg)+\cdots,}
\end{equation}
with
\begin{eqnarray}
a^{(1)}_{n,m}(\mu_i,\mu_j)=\frac{n^2(Un-\mu_i+\mu_j)}{(U(n-m-1)-\mu_i+\mu_j)(U(n-m)-\mu_i+\mu_j)},\\
a^{(2)}_{n,m}(\mu_i,\mu_j)=\frac{(n+1)^2(U(n+1)-\mu_i+\mu_j)}{(U(n-m+1)-\mu_i+\mu_n)(U(n-m)-\mu_i+\mu_n)},
\end{eqnarray}
where we show only those terms which will be important to our analysis. We have chosen to represent the double poles implicit in \eref{eq34} as derivatives of first-order poles. This representation simplifies the process of analytic continuation which must be taken in order to go from the Matsubara frequencies to the real frequency domain. Such an analytic continuation can be expressed by the transformation $\rmi\omega_{l}\rightarrow\omega\pm \rmi\eta$, with $\eta\rightarrow 0^+$. Following \cite{abrikosov1963methods}, the density of states can be obtained by using $\rho_i(\omega)=-\frac{1}{\pi} \mbox{Im}(G_i(\omega))$. Hence, in the second-order hopping expansion it gives
\begin{equation}
\fl
\eqalign{\rho_i(\omega)=&\frac{1}{Z_0(\mu_i)}\sum_{n=0}^\infty \rme^{-\beta f_n(\mu_i)}\Big[n\delta(\omega+\mu_i-U(n-1))+(n+1)\delta(\omega+\mu_i-Un)\Big]\\
&+
\sum_jJ_{ij}J_{ji}\sum_{n,m=0}^\infty \Bigg(\frac{\rme^{-\beta f_n(\mu_i)}}{Z_0(\mu_i)}\Bigg)^2\frac{\rme^{-\beta f_m(\mu_j)}}{Z_0(\mu_j)}
\\
&\quad\times\frac{\partial}{\partial\mu_i}\Bigg(
a^{(1)}_{n,m}(\mu_i,\mu_j)\delta(\omega+\mu_i-U(n-1))-
a^{(2)}_{n,m}(\mu_i,\mu_j)\delta(\omega+\mu_i-Un)\Bigg)\\
&+\cdots.}
\end{equation}
By taking the disorder ensemble average of the density of states we  get
\begin{equation}
\fl
\eqalign{\overline{\rho}(\omega,\mu)=&\sum_{n=0}^\infty n\frac{\rme^{-\beta f_n(U(n-1)-\omega)}p(U(n-1)-\omega-\mu)}{Z_0(U(n-1)-\omega)}
\\&+
\sum_{n=0}^\infty (n+1)\frac{\rme^{-\beta f_n(Un-\omega)}p(Un-\omega-\mu)}{Z_0(Un-\omega)}
\\&+
\sum_jJ_{ij}J_{ji}\sum_{n=0}^\infty n^2\frac{\rme^{-2\beta f_n(U(n-1)-\omega)}}{Z_0^2(U(n-1)-\omega)}
\xi(\mu,\omega)\frac{\partial(p(\mu_i-\mu))}{\partial\mu_i}\Bigg\vert_{\mu_i=U(n-1)-\omega}
\\&-
\sum_jJ_{ij}J_{ji}\sum_{n=0}^\infty (n+1)^2\frac{\rme^{-2\beta f_n(Un-\omega)}}{Z_0^2(Un-\omega)}
\xi(\mu,\omega)\frac{\partial(p(\mu_i-\mu))}{\partial\mu_i}\Bigg\vert_{\mu_i=Un-~\omega}\\
&+\cdots,}\label{eq39}
\end{equation}
where 
\begin{equation}
\fl\xi(\mu,\omega)=\int^{\mu+\infty}_{\mu-\infty}\rmd \mu_j\frac{p(\mu_j-\mu)}{Z_0(\mu_j)}\sum_{m=0}^\infty\frac{m+1}{\omega+\mu_j-Um}(\rme^{-\beta f_{m+1}(\mu_j)}-\rme^{-\beta f_m(\mu_j)}),
\label{eq40}
\end{equation}
which, in the case where $\omega=0$, is essentially the same integration as \eref{eq30}.

The derivatives that appear in the second-order correction of \eref{eq39} are a product of our perturbation theory. In the exact solution, the Green's function should only present simple poles in Matsubara space, as is demonstrate, for instance, in \cite[Chapter 9]{fetter2012quantum} or \cite[Chapter 3]{mahan2013many}. Therefore, these anomalies in the expression of the density of states are related to the double poles implicit in \eref{eq34}. In order to deal with such a problem, one must renormalize the location of these poles. For that reason, we propose the following transformations
\begin{equation}
\eqalign{
\gamma^-_n=U(n-1)-\omega-\mu\quad&\rightarrow\quad \Omega^-_n+\lambda(\gamma^-_n-\Omega^-_n),\\
\gamma^+_n=Un-\omega-\mu &\rightarrow\quad\Omega^+_n+\lambda(\gamma^+_n-\Omega^+_n),\\
 J_{ij}J_{ji} \quad&\rightarrow\quad \lambda J_{ij}J_{ji},}\label{eq41}
\end{equation}
where the initial situation is recovered by setting $\lambda=1$. The idea is to use these transformation to expand the density of states in $\lambda$ and then determine the values of the renormalized frequencies that eliminate the derivatives that appear in \eref{eq39}.

In the Poincaré-Lindstedt method, the renormalization of the frequency aims to eliminate secular terms in a response that should be periodic \cite{bender1999advanced,pelster2003high,vidanovic2011nonlinear,al2013geometric}. If we take a Fourier transform in time from the Poincaré-Lindstedt method, purely periodic terms will appear as simple poles, while secular terms will appear as higher order poles. Analogously, we can understand the delta functions as the imaginary part of the simple poles and its derivatives as the imaginary part of higher order poles. In this fashion, we can interpret the renormalization of the poles in the Green's function approach as the Poincaré-Lindstedt method in the Fourier space. Therefore, the renormalization of the delta functions is equivalent to the previous method considering only the imaginary parts. 

Using the transformations of \eref{eq41}, the first-order in the $\lambda$ expansion for the density of states gives
\begin{equation}
\eqalign{\overline{\rho}(\omega,\mu)=&\sum_{n=0}^\infty n\frac{\rme^{-\beta f_n(U(n-1)-\omega)}p(\Omega^-)}{Z_0(U(n-1)-\omega)}
\\&+
\sum_{n=0}^\infty (n+1)\frac{\rme^{-\beta f_n(Un-\omega)}p(\Omega^+)}{Z_0(Un-\omega)}
\\&+
\lambda\sum_{n=0}^\infty n\frac{\rme^{-\beta f_n(U(n-1)-\omega)}(\gamma^-_n-\Omega^-)}{Z_0(U(n-1)-\omega)}p^\prime(\Omega^-)
\\&+
\lambda\sum_jJ_{ij}J_{ji}\sum_{n=0}^\infty n^2\frac{\rme^{-2\beta f_n(U(n-1)-\omega)}}{Z_0^2(U(n-1)-\omega)}
\xi(\mu,\omega)p^\prime(\Omega^-)
\\&+
\lambda\sum_{n=0}^\infty (n+1)\frac{\rme^{-\beta f_n(Un-\omega)}(\gamma^+_n-\Omega^+)}{Z_0(Un-\omega)}p^\prime(\Omega^+)
\\&-
\lambda\sum_jJ_{ij}J_{ji}\sum_{n=0}^\infty (n+1)^2\frac{\rme^{-2\beta f_n(Un-\omega)}}{Z_0^2(Un-\omega)}
\xi(\mu,\omega)p^\prime(\Omega^+)
\\&+\cdots.}\label{eq42}
\end{equation}
Setting $\lambda=1$, we now choose the values of $\Omega^-_n$ and $\Omega^+_n$ such that they cancel the derivatives of the disorder distribution in \eref{eq42} thus finding the following expressions
\begin{equation}
\Omega^-_n=U(n-1)-\omega-\mu+n\frac{\rme^{-\beta f_n(U(n-1)-\omega)}}{Z_0(U(n-1)-\omega)}\sum_jJ_{ij}J_{ji}\xi(\mu,\omega),
\end{equation}
\begin{equation}
\Omega^+_n=Un-\omega-\mu-(n+1)\frac{\rme^{-\beta f_n(Un-\omega)}}{Z_0(Un-\omega)}\sum_jJ_{ij}J_{ji}\xi(\mu,\omega).
\end{equation}

The result of this process is that the relevant correction to the density of states in the second hopping order expansion consists of a shift in the argument of the disorder distribution function, as can be observed in the equation below
\begin{equation}
\fl
\eqalign{
\overline{\rho}(\omega,\mu)=&\sum_{n=0}^\infty n\frac{\rme^{-\beta f_n(U(n-1)-\omega)}}{Z_0(U(n-1)-\omega)}\\
&\qquad\times
p\Big(U(n-1)-\omega-\mu+n\frac{\rme^{-\beta f_n(U(n-1)-\omega)}}{Z_0(U(n-1)-\omega)}J^2z\xi(\mu,\omega)\Big)
\\& +
\sum_{n=0}^\infty (n+1)\frac{\rme^{-\beta f_n(Un-\omega)}}{Z_0(Un-\omega)}
\\
&\qquad\times
p\Big(Un-\omega-\mu-(n+1)\frac{\rme^{-\beta f_n(Un-\omega)}}{Z_0(Un-\omega)}J^2z\xi(\mu,\omega)\Big)+\cdots,}\label{eq45}
\end{equation}
where we have considered hopping processes only between first neighbouring sites. Note that \eref{eq45} constitutes a resummation of \eref{eq39} and consequently both equations are equivalent up to second order in $J$. With this result, we have therefore computed the corrections due to the hopping of particles to the expression of the density of states calculated in \cite{krutitsky2006mean}.

Now the task is to solve the integral of \eref{eq40}. By choosing a uniform disorder distribution of the form of \eref{eq31} such an integration is reduced to
\begin{equation}
\fl\xi(\mu,\omega)=\frac{1}{\Delta}\int^{\mu+\frac{\Delta}{2}}_{\mu-\frac{\Delta}{2}}\rmd\mu_j\frac{1}{Z_0(\mu_j)}\sum_{m=0}^\infty\frac{m+1}{\omega+\mu_j-Um}(\rme^{-\beta f_{m+1}(\mu_j)}-\rme^{-\beta f_m(\mu_j)}).
\label{eq46}
\end{equation}
\begin{figure}[!]
\centering
\subfloat[]{\includegraphics[width=0.48\textwidth]{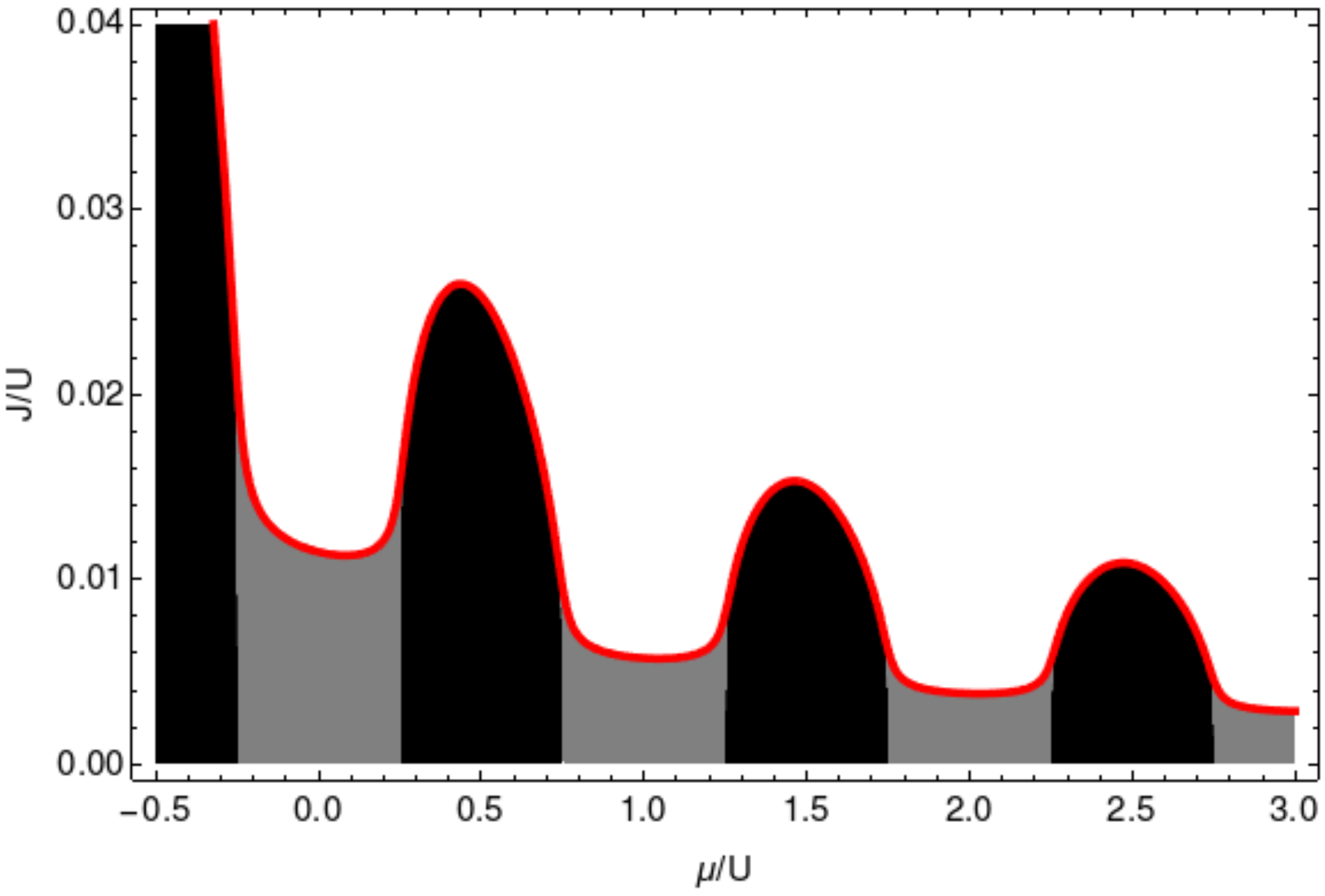}}
\hfill
\subfloat[]{\includegraphics[width=0.48\textwidth]{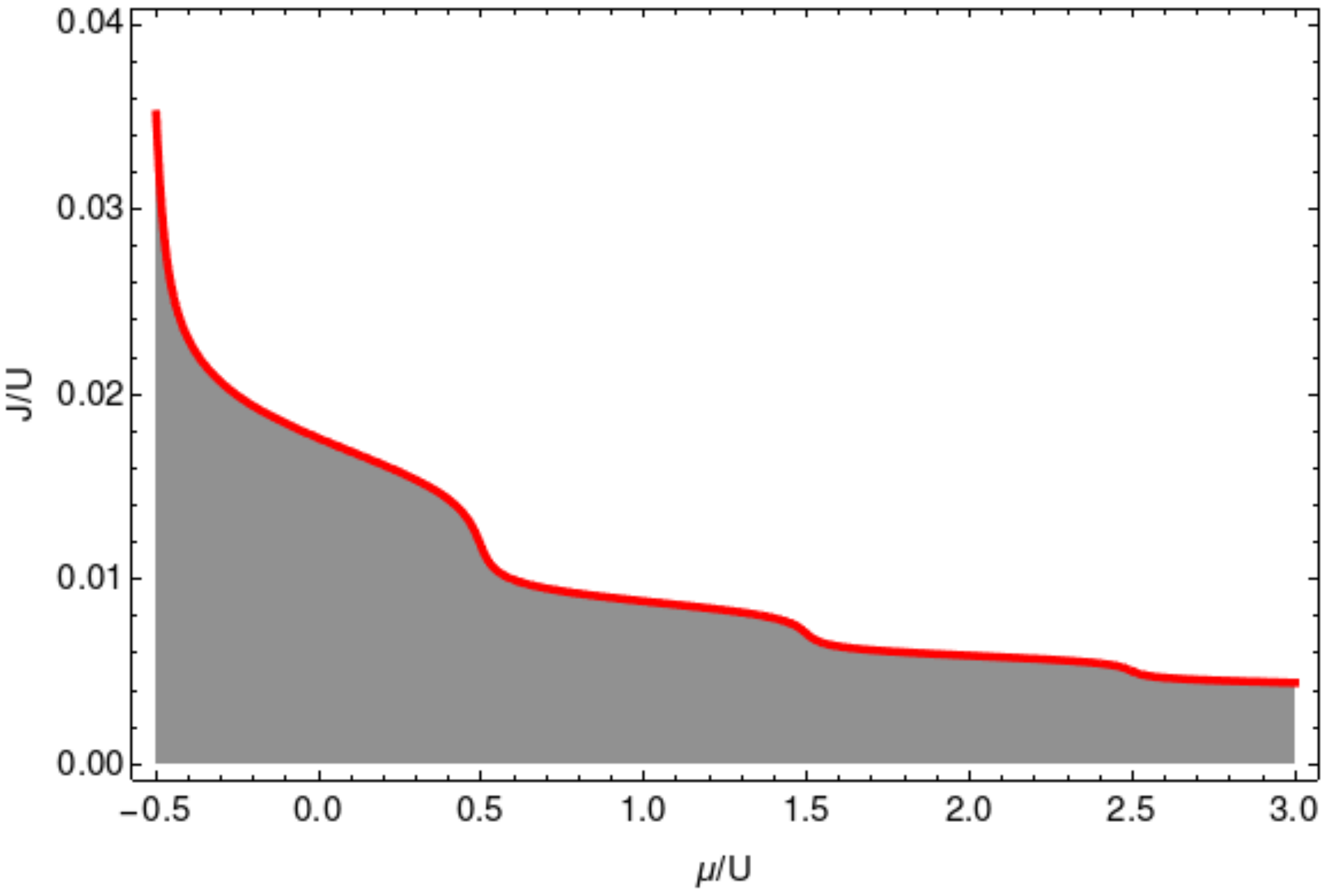}}
\caption{Density plot of the single-particle density of states obtained from \eref{eq45} for $k_BT/U=0.01$ and $z=6$. The black regions correspond to the Mott-insulator phase, where $\overline{\rho}(0,\mu)=0$, and the grey region corresponds to the Bose-glass phase, where $\overline{\rho}(0,\mu)\neq 0$. The red line indicates the resummed Green's function result for the phase boundary to the superfluid phase as shown in \fref{fig1}, which corresponds to the white region of the figures. The disorder strength is fixed with $\Delta/U=0.5$ in (a) and $\Delta/U=1$ in (b).}\label{fig2}
\end{figure}

As explained earlier, the Mott insulator and the Bose glass can be distinguished from one another by the fact that, in the case of $\omega=0$, the single-particle density of states is zero in the Mott phase and finite in the Bose-glass phase. The result of \eref{eq45} allows us to construct a phase diagram by distinguishing on a density plot the regions below the superfluid line \eref{eq32} where $\overline{\rho}(0,\mu)$ vanishes from the ones where it is finite, which can be observed in \fref{fig2} for homogeneous disorder. As argued in \cite{fisher1989boson}, in the case where $\Delta<U$, there should always be a region, namely $\mu\in [U(n-1)+\Delta/2, Un-\Delta/2]$ for $J=0$, where the average number of particles that minimizes the local energy $f_n(\mu_i)$ given in \eref{eq22} becomes fixed at an integer value, which represents the Mott insulating state. When $J$ is made slightly positive, this situation is sustained and both Mott insulator and Bose glass are guaranteed to appear at different regions of the insulating part of the phase diagram. As can be seen in \fref{fig2}(a), the Bose-glass phase, represented by the grey regions, emerges between the Mott lobes which correspond to the black regions. In the case where $\Delta\geq U$, the average number of particles per site never sticks to an integer value and the Bose-glass phase suppresses the Mott states dominating the insulating part of the phase diagram, which is shown in \fref{fig2}(b). In the two cases, above the red line, only the superfluid phase exists. The result shown in \fref{fig2} demonstrates that the single-particle density of states distinguishes unambiguously the Mott insulator from the Bose-glass state, however it gives no detail about the exact phase boundary between the two. In the situation where the three phases coexist, our perturbation theory should be valid to determine the such a phase boundary. This calculation will now be presented.
\section{Mott insulator to Bose glass phase boundary}\label{S5}

\begin{figure}[!]
\centering
\includegraphics[width=0.55\textwidth]{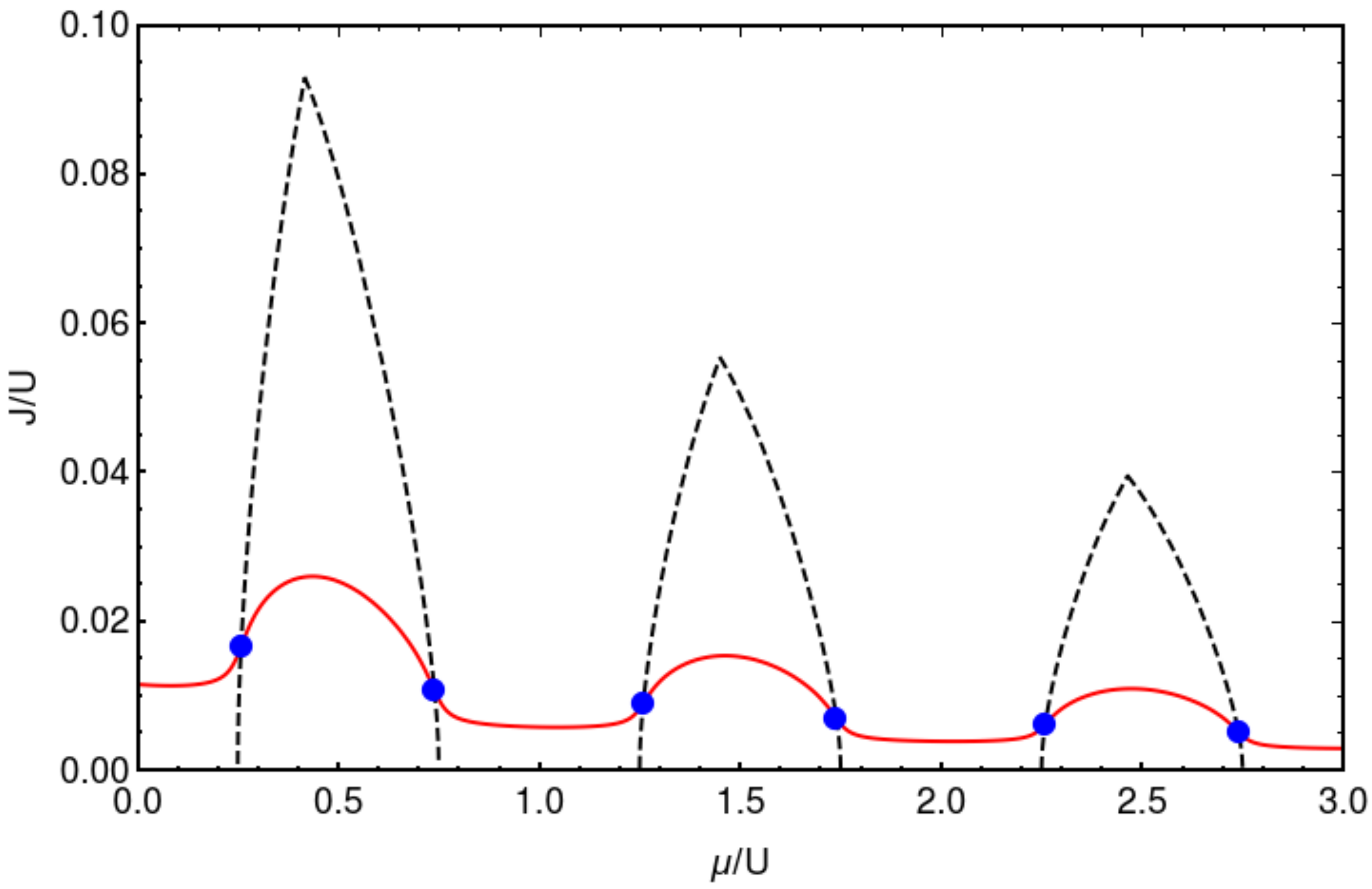}
\caption{Phase boundaries for the case of $k_BT/U=0.01$, $z=6$, and $\Delta/U=0.5$. The continuous red line corresponds to \eref{eq32}, the result of the resummed Green's function approach, and the dashed black lines indicate \eref{eq49}, the phase boundary between Mott insulator and Bose glass obtained based on the result for the single-particle density of states. The blue dots represent the tri-critical points where the three phases coexist.}\label{fig3}
\end{figure}

We can calculate the analytical phase boundary between Mott insulator and Bose glass by analysing the distributions in \eref{eq45}. For the uniform disorder distribution of \eref{eq31}, the Bose-glass region, where $\overline{\rho}(0,\mu)\neq 0$, is limited by the following inequalities
\begin{equation}\label{eq47}
\eqalign{
-\frac{\Delta}{2}\leq\gamma_n^{-}+z\Lambda^{-}_nJ^2\leq\frac{\Delta}{2},
\\
-\frac{\Delta}{2}\leq\gamma_n^{+}-z\Lambda^{+}_nJ^2\leq\frac{\Delta}{2},}
\end{equation}
where we have defined
\begin{equation}
\eqalign{
\gamma_n^{-}\quad &=\quad U(n-1)-\mu,
\\
\gamma_n^{+}\quad &=\quad Un-\mu,
\\
\Lambda^{-}_n\quad &=\quad n\frac{\rme^{-\beta f_n(U(n-1))}\xi(\mu,0)}{Z_0(U(n-1))},
\\
\Lambda^{+}_n\quad &=\quad (n+1)\frac{\rme^{-\beta f_n(Un)}\xi(\mu,0)}{Z_0(Un)}.}
\end{equation}
As explained in the last section, for sufficiently weak disorder, $\Delta<U$, the Mott states should always appear in the insulating part of the phase diagram. Therefore, based on \eref{eq47}, the Mott lobes would correspond to the regions satisfying the following condition to the hopping parameter
\begin{equation}
\Bigg(\frac{-\frac{\Delta}{2}-\gamma_n^{-}}{z\Lambda^{-}_n}\Bigg)^{\case{1}{2}}\leq J\leq\Bigg(\frac{-\frac{\Delta}{2}+\gamma_n^{+}}{z\Lambda^{+}_n}\Bigg)^{\case{1}{2}}.
\label{eq49}
\end{equation}
What stands out in this result is the fact that the location of any Mott lobe can be obtained by plugging in the above equation the integer $n$ which minimizes the local energy. The phase boundaries of \eref{eq32} and \eref{eq49} are shown in \fref{fig3} for a uniform disorder distribution.

According to our analysis, the region inside the dashed black lines corresponds to the Mott lobes, where $\overline{\rho}(0,\mu)=0$, and outside of this region, where $\overline{\rho}(0,\mu)\neq 0$, we would have the Bose-glass phase. With our approximation, which corresponds to calculating hopping-dependent corrections to the density of states analysed in \cite{krutitsky2006mean}, we have found an equation for the phase boundary between the two insulating phases valid for small values of $J$. Our finding shows that our Green's function approach is capable of going beyond mean-field theory for the prediction to the phase boundary between the insulating phases, demonstrating that the single-particle density of states still serves to unambiguously distinguish the Bose-glass and Mott-insulator phases at finite temperatures and when the tunnelling energy is made slightly positive. However, above the continuous red line, which corresponds to the result of \eref{eq32}, only the superfluid phase should exist. As we can see, the Mott lobes of \eref{eq49} trespass this limit, allowing a direct transition between Mott insulator and superfluid phase. It has been analytically proved that such a transition should not occur in the presence of any bounded disorder \cite{pollet2009absence}. Our result may be explained by the fact that we have considered only the first relevant hopping correction to the single-particle density of states. We expect that by considering higher order approximations to this quantity, the triple critical points, represented by blue dots located at both sides of each Mott lobe, would join below the red curve, such that the Bose glass would always intervene between the Mott-insulator and superfluid phases, never allowing a direct transition between the latter two. Further analysis of the energy behaviour inside that region could also lead to a determination of an accurate phase boundary. An effective-action approach, similar to the one of \cite{dos2009quantum,bradlyn2009effective,dos2011ginzburg}, should also be of use in this respect, where an Edwards-Anderson like order parameter, such as the one suggested in \cite{graham2009order} and used in \cite{thomson2016measuring,khellil2016hartree,khellil2016analytical,khellil2017dirty}, could be applied to identify the Bose-glass phase, allowing also to obtain information on the nature its collective excitations. The investigation of these questions characterizes a natural progression of this work.

We now turn our attention to the comparison between our results and numerical data from the literature.
\section{Comparison with numerical results}\label{S6}
By considering the results of \eref{eq49} only below the curve of \eref{eq32}, we can locate the region corresponding to any Mott lobe in the phase diagram. This makes a direct comparison between our results and numerical data from the literature possible. To this end, we first consider our prediction of the phase boundary in the zero-temperature limit, which corresponds to $\beta\rightarrow\infty$. This results in
\begin{equation}
J=\frac{1}{z\xi_0(0,\mu)},
\label{eq50}
\end{equation}
for the phase boundary with the superfluid phase, and 
\begin{equation}
\Bigg(\frac{-\frac{\Delta}{2}-\gamma_{n_0}^{-}}{zn_0\xi_0(0,\mu)}\Bigg)^{\case{1}{2}}\leq J\leq\Bigg(\frac{-\frac{\Delta}{2}+\gamma_{n_0}^{+}}{z(n_0+1)\xi_0(0,\mu)}\Bigg)^{\case{1}{2}},
\label{eq51}
\end{equation}
for the Mott insulator to Bose glass phase boundary, where we have defined
\begin{equation}
\fl\eqalign{\xi_0(0,\mu)&=\lim_{\beta\to\infty}\xi(0,\mu)\\
&=\frac{n_0}{\Delta}\ln\Bigg(\frac{\mu+\frac{\Delta}{2}-U(n_0-1)}{\mu-\frac{\Delta}{2}-U(n_0-1)}\Bigg)+\frac{(n_0+1)}{\Delta}\ln\Bigg(\frac{\mu-\frac{\Delta}{2}-Un_0}{\mu+\frac{\Delta}{2}-Un_0}\Bigg),}
\end{equation}
and $n_0$ minimizes the energy inside each Mott lobe. 

\begin{figure}[!]
\subfloat[]{\includegraphics[width=0.33\textwidth]{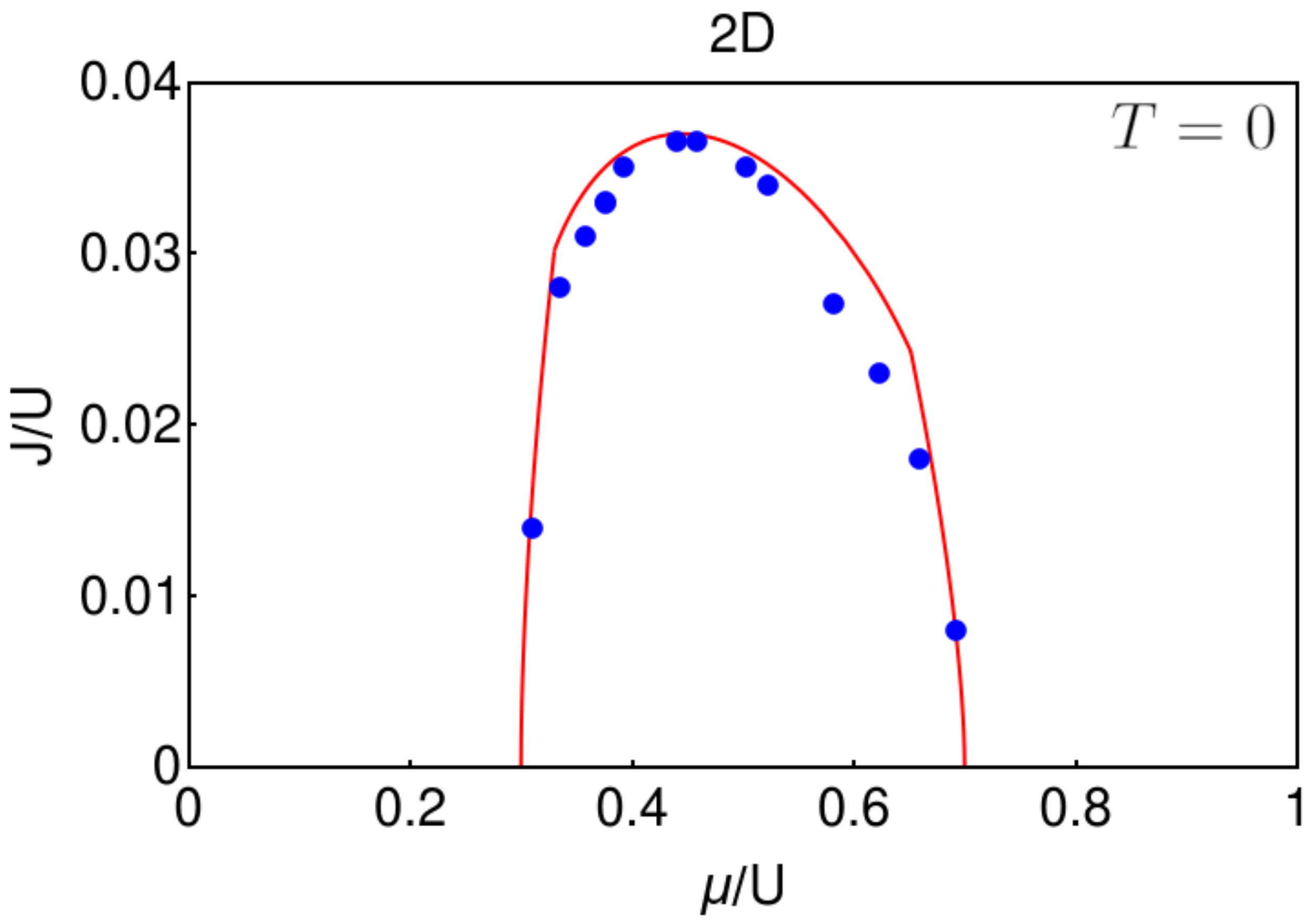}}
\subfloat[]{\includegraphics[width=0.33\textwidth]{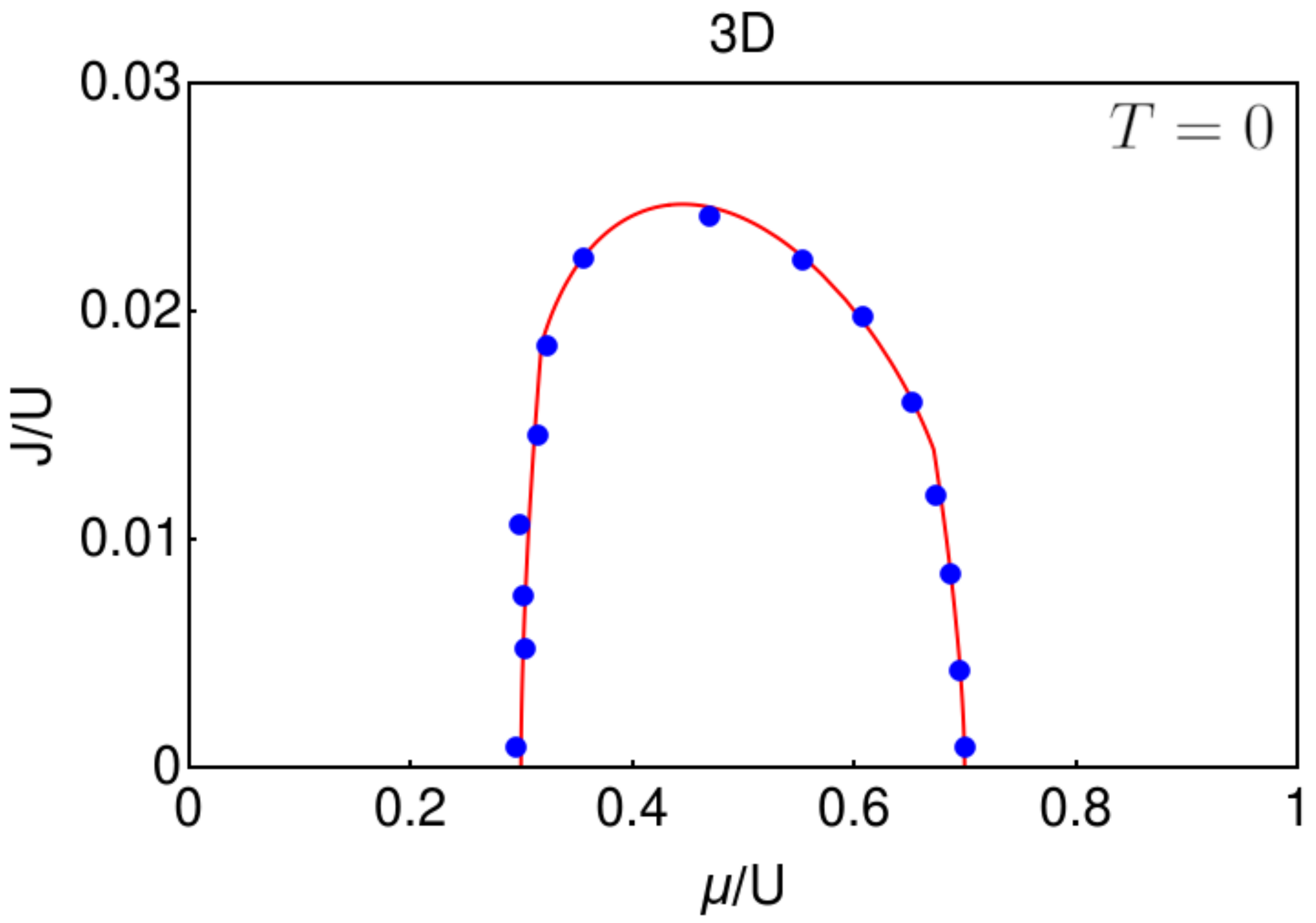}}
\subfloat[]{\includegraphics[width=0.33\textwidth]{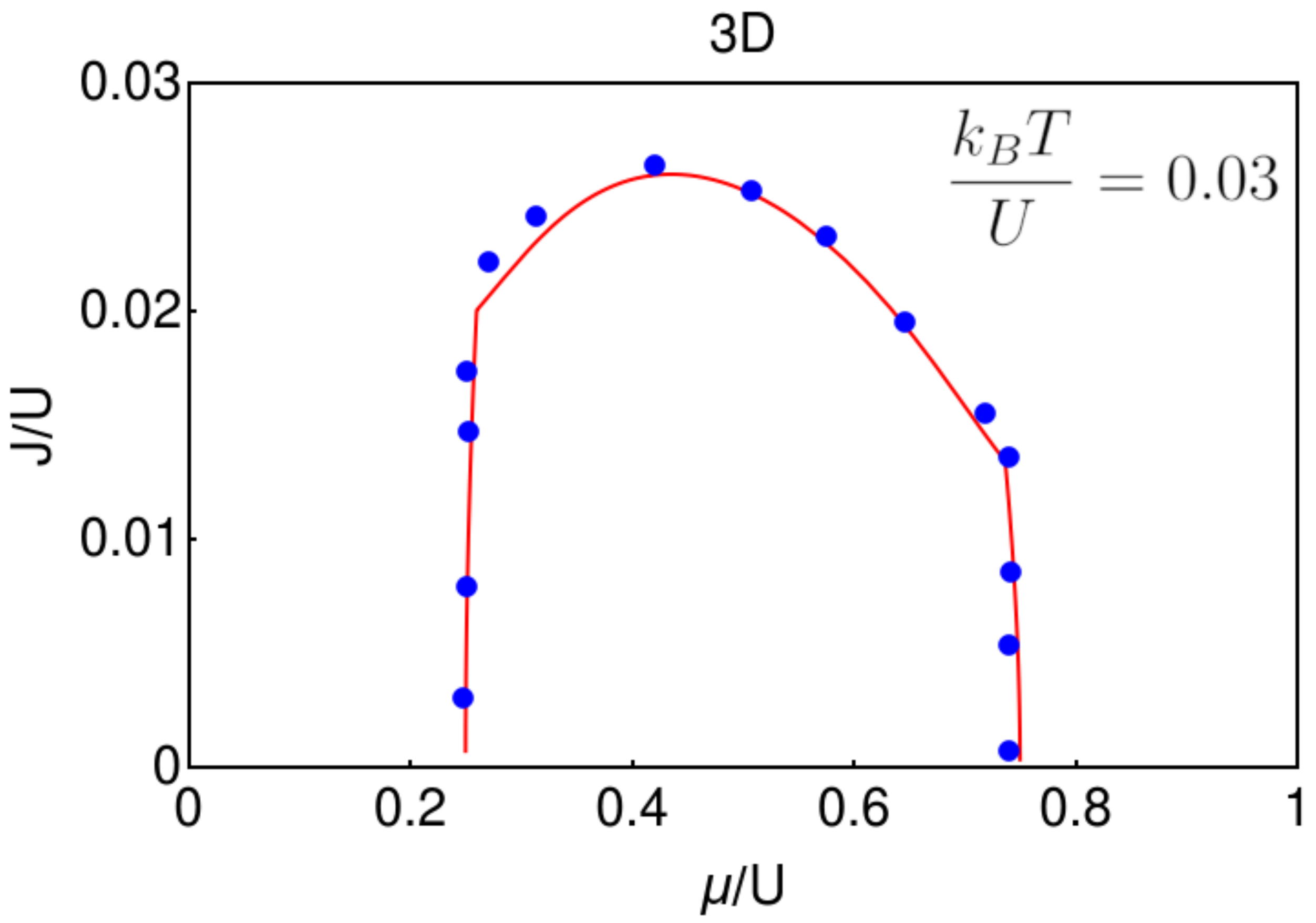}}
\caption{Comparison between theoretical and numerical results for the first Mott lobe in the disordered case for two and three dimensions at both zero and finite temperatures. The continuous red line corresponds to \eref{eq50} and \eref{eq51}, the results of our Green's function approach, while the blue dots indicate the numerical predictions of \cite{thomson2016measuring} for the $2D$ case and \cite{bissbort2009stochastic,bissbort2010stochastic} for the $3D$ case. Figures (a) and (b) show the $2D$ and $3D$ zero-temperature phase boundary with $\Delta/U=0.6$ and $z=4$ and $z=6$, respectively. Figure (c) presents the finite temperature $3D$ phase boundary with $\Delta/U=0.5$, $z=6$, and $k_BT/U=0.03$.}\label{fig4}
\end{figure}

Using the above expressions, we are now able to compare our result with the numerical data of \cite{thomson2016measuring}, where an Edwards-Anderson order parameter was used in order to characterize the Bose-glass phase for a $2D$ square lattice at zero temperature. For the $3D$ case, we use the data from \cite{bissbort2009stochastic,bissbort2010stochastic} for zero and finite temperatures, respectively. Such a comparison can be observed in \fref{fig4}.

As is demonstrated, our results compare quite well with the numerical data from the literature. In \fref{fig4}(a), the numerical points lie mostly inside the red continuous line indicating that our prediction overestimates the first Mott lobe in the $2D$ case. However, in \ref{fig4}(b) and \ref{fig4}(c), we see a remarkable agreement of our prediction with numerical data for the $3D$ phase boundary for both zero and finite temperatures. We notice in \fref{fig4}(c) that the introduction of temperature leads to a smoother curve for the phase boundary to the superfluid phase, characterized by \eref{eq32}. As a consequence, in the points where this solution meets the Mott insulator to Bose glass phase boundary of \eref{eq49}, kinks emerge, locating exactly the tri-critical points where the three phases coexist. Despite the discrepancy in the comparison for the $2D$ case, the relative deviation between our prediction and the numerical data for the tip of the Mott lobe in the three cases represents an error of less than $2\%$. Therefore, we can conclude that our results are in significant accordance with the numerical calculations.

\section{Summary and Conclusions}\label{S7}

We have presented an analytic approach for perturbatively calculating the finite-temperature Green's function of the disordered Bose-Hubbard model as well as the analysis of this quantity in order to provide a distinction for the three possible ground states of the system. By summing up a subset of the contributions in the hopping expansion of the $2$-point Green's function we were able to reproduce the results of mean-field theory for the phase boundary between the superfluid and insulating phases obtained in \cite{krutitsky2006mean}. A renormalization method was employed allowing us to compute the first relevant correction to single-particle density of states due to the hopping of particles, which made it possible to construct a phase diagram with unambiguous distinction between Mott insulator and Bose-glass, thus confirming that this quantity still serves to differentiate both phases even for slightly positive values of the kinetic energy. Our results compared well with the numerical results of \cite{thomson2016measuring} for the $2D$ zero-temperature phase boundary and show noticeable agreement with the $3D$ numerical results of \cite{bissbort2009stochastic,bissbort2010stochastic} both for zero and finite temperatures, providing an error of less than $2\%$ for the first Mott lobe tip. However, our result for the phase diagram predicts the possibility of a direct transition between Mott-insulator and superfluid phases, which should not occur in the presence of any bounded disorder, as was proven in \cite{pollet2009absence}. Despite this limitation, it was made clear that this approach is able of going beyond mean-field theory for the distinction of the two insulating phases and we expect that the inclusion of higher order corrections in the hopping expansion of the density of states should lead to a more accurate prediction to the superfluid to Bose-glass phase transition. Further studies regarding these questions should be the subject of future investigation.

\ack
This study was financed in part by CAPES (Coordenação de Aperfeiçoamento de Pessoal de Nível Superior, Improvement Coordination of Higher Level Personnel) – Brasil – Finance Code 001. We thank the binational project between CAPES and DAAD (Deutscher Akademischer Austauschdienst, German Academic Exchange Service). Furthermore, A. P. acknowledges financial support by the Deutsche Forschungsgemeinschaft (DFG, German Research Foundation) via the Collaborative Research Center SFB/TR185 (Project No. 277625399). F. E. A. S. thanks CNPq (Conselho Nacional de Desenvolvimento Científico e Tecnológico, National Council for Scientific and Technological Development) for support through Bolsa de produtividade em Pesquisa (Research productivity scholarship) Grant No. 305586/2017-3.

\appendix
\renewcommand{\theequation}{A.\arabic{equation}}
\setcounter{equation}{0}

\section*{References}

\bibliographystyle{unsrt}
\bibliography{references}

\begin{thebibliography}{10}

\bibitem{fisher1989boson}
Matthew~P.A. Fisher, Peter~B Weichman, G~Grinstein, and Daniel~S Fisher.
\newblock Boson localization and the superfluid-insulator transition.
\newblock {\em Physical Review B}, 40(1):546, 1989.

\bibitem{bloch2005ultracold}
Immanuel Bloch.
\newblock Ultracold quantum gases in optical lattices.
\newblock {\em Nature Physics}, 1(1):23, 2005.

\bibitem{lye2005bose}
J.E. Lye, L~Fallani, Michele Modugno, DS~Wiersma, C~Fort, and Massimo Inguscio.
\newblock Bose-einstein condensate in a random potential.
\newblock {\em Physical review letters}, 95(7):070401, 2005.

\bibitem{fallani2007ultracold}
L~Fallani, JE~Lye, V~Guarrera, C~Fort, and M~Inguscio.
\newblock Ultracold atoms in a disordered crystal of light: Towards a bose
  glass.
\newblock {\em Physical review letters}, 98(13):130404, 2007.

\bibitem{wang2004disordered}
Daw-Wei Wang, Mikhail~D Lukin, and Eugene Demler.
\newblock Disordered bose-einstein condensates in quasi-one-dimensional
  magnetic microtraps.
\newblock {\em Physical review letters}, 92(7):076802, 2004.

\bibitem{schumm2005atom}
Thorsten Schumm, J{\'e}r{\^o}me Est{\`e}ve, C~Figl, J-B Trebbia, C~Aussibal,
  Hai Nguyen, Dominique Mailly, Isabelle Bouchoule, Christoph~I Westbrook, and
  Alain Aspect.
\newblock Atom chips in the real world: the effects of wire corrugation.
\newblock {\em The European Physical Journal D-Atomic, Molecular, Optical and
  Plasma Physics}, 32(2):171--180, 2005.

\bibitem{clement2005suppression}
David Cl{\'e}ment, Andres~F Varon, Mathilde Hugbart, Jocelyn~A Retter, Philippe
  Bouyer, Laurent Sanchez-Palencia, Dimitri~M Gangardt, Georgy~V Shlyapnikov,
  and Alain Aspect.
\newblock Suppression of transport of an interacting elongated bose-einstein
  condensate in a random potential.
\newblock {\em Physical Review Letters}, 95(17):170409, 2005.

\bibitem{billy2008direct}
Juliette Billy, Vincent Josse, Zhanchun Zuo, Alain Bernard, Ben Hambrecht,
  Pierre Lugan, David Cl{\'e}ment, Laurent Sanchez-Palencia, Philippe Bouyer,
  and Alain Aspect.
\newblock Direct observation of anderson localization of matter waves in a
  controlled disorder.
\newblock {\em Nature}, 453(7197):891--894, 2008.

\bibitem{zhou2010construction}
SQ~Zhou and DM~Ceperley.
\newblock Construction of localized wave functions for a disordered optical
  lattice and analysis of the resulting hubbard model parameters.
\newblock {\em Physical Review A}, 81(1):013402, 2010.

\bibitem{bakr2009quantum}
Waseem~S Bakr, Jonathon~I Gillen, Amy Peng, Simon F{\"o}lling, and Markus
  Greiner.
\newblock A quantum gas microscope for detecting single atoms in a
  hubbard-regime optical lattice.
\newblock {\em Nature}, 462(7269):74--77, 2009.

\bibitem{bender2018customizing}
Nicholas Bender, Hasan Y{\i}lmaz, Yaron Bromberg, and Hui Cao.
\newblock Customizing speckle intensity statistics.
\newblock {\em Optica}, 5(5):595--600, 2018.

\bibitem{greiner2002quantum}
Markus Greiner, Olaf Mandel, Tilman Esslinger, Theodor~W H{\"a}nsch, and
  Immanuel Bloch.
\newblock Quantum phase transition from a superfluid to a mott insulator in a
  gas of ultracold atoms.
\newblock {\em Nature}, 415(6867):39, 2002.

\bibitem{meldgin2016probing}
Carolyn Meldgin, Ushnish Ray, Philip Russ, David Chen, David~M Ceperley, and
  Brian DeMarco.
\newblock Probing the bose glass--superfluid transition using quantum quenches
  of disorder.
\newblock {\em Nature Physics}, 12(7):646, 2016.

\bibitem{astrakharchik2002superfluidity}
GE~Astrakharchik, J~Boronat, J~Casulleras, and S~Giorgini.
\newblock Superfluidity versus bose-einstein condensation in a bose gas with
  disorder.
\newblock {\em Physical Review A}, 66(2):023603, 2002.

\bibitem{capogrosso2007phase}
B~Capogrosso-Sansone, NV~Prokof’Ev, and BV~Svistunov.
\newblock Phase diagram and thermodynamics of the three-dimensional
  bose-hubbard model.
\newblock {\em Physical Review B}, 75(13):134302, 2007.

\bibitem{meier2012quantum}
Hannes Meier and Mats Wallin.
\newblock Quantum critical dynamics simulation of dirty boson systems.
\newblock {\em Physical Review Letters}, 108(5):055701, 2012.

\bibitem{zhang2015equilibrium}
Chao Zhang, Arghavan Safavi-Naini, and Barbara Capogrosso-Sansone.
\newblock Equilibrium phases of two-dimensional bosons in quasiperiodic
  lattices.
\newblock {\em Physical Review A}, 91(3):031604, 2015.

\bibitem{ng2015quantum}
Ray Ng and Erik~S S{\o}rensen.
\newblock Quantum critical scaling of dirty bosons in two dimensions.
\newblock {\em Physical Review Letters}, 114(25):255701, 2015.

\bibitem{de2018properties}
Bruno~R de~Abreu, Ushnish Ray, Silvio~A Vitiello, and David~M Ceperley.
\newblock Properties of the superfluid in the disordered bose-hubbard model.
\newblock {\em Physical Review A}, 98(2):023628, 2018.

\bibitem{bissbort2009stochastic}
Ulf Bissbort and Walter Hofstetter.
\newblock Stochastic mean-field theory for the disordered bose-hubbard model.
\newblock {\em EPL (Europhysics Letters)}, 86(5):50007, 2009.

\bibitem{bissbort2010stochastic}
Ulf Bissbort, Ronny Thomale, and Walter Hofstetter.
\newblock Stochastic mean-field theory: Method and application to the
  disordered bose-hubbard model at finite temperature and speckle disorder.
\newblock {\em Physical Review A}, 81(6):063643, 2010.

\bibitem{thomson2016measuring}
Steven~J Thomson, Liam~S Walker, Tiffany~L Harte, and Graham~D Bruce.
\newblock Measuring the edwards-anderson order parameter of the bose glass: A
  quantum gas microscope approach.
\newblock {\em Physical Review A}, 94(5):051601, 2016.

\bibitem{krutitsky2006mean}
KV~Krutitsky, A~Pelster, and R~Graham.
\newblock Mean-field phase diagram of disordered bosons in a lattice at nonzero
  temperature.
\newblock {\em New Journal of Physics}, 8(9):187, 2006.

\bibitem{buonsante2007mean}
P~Buonsante, V~Penna, A~Vezzani, and PB~Blakie.
\newblock Mean-field phase diagram of cold lattice bosons in disordered
  potentials.
\newblock {\em Physical Review A}, 76(1):011602, 2007.

\bibitem{pisarski2011application}
P~Pisarski, RM~Jones, and RJ~Gooding.
\newblock Application of a multisite mean-field theory to the disordered
  bose-hubbard model.
\newblock {\em Physical Review A}, 83(5):053608, 2011.

\bibitem{dos2009quantum}
FEA Dos~Santos and A~Pelster.
\newblock Quantum phase diagram of bosons in optical lattices.
\newblock {\em Physical Review A}, 79(1):013614, 2009.

\bibitem{bradlyn2009effective}
Barry Bradlyn, Francisco Ednilson~A Dos~Santos, and Axel Pelster.
\newblock Effective action approach for quantum phase transitions in bosonic
  lattices.
\newblock {\em Physical Review A}, 79(1):013615, 2009.

\bibitem{dos2011ginzburg}
Francisco Ednilson~Alves dos Santos.
\newblock {\em Ginzburg-Landau theory for bosonic gases in optical lattices}.
\newblock PhD thesis, Freie Universit{\"a}t Berlin, 2011.

\bibitem{ohliger2013green}
Matthias Ohliger and Axel Pelster.
\newblock Green's function approach to the bose-hubbard model.
\newblock {\em World Journal of Condensed Matter Physics}, 3(2):125--130, 2013.

\bibitem{yao2014critical}
Zhiyuan Yao, Karine~PC da~Costa, Mikhail Kiselev, and Nikolay Prokof’ev.
\newblock Critical exponents of the superfluid--bose-glass transition in three
  dimensions.
\newblock {\em Physical review letters}, 112(22):225301, 2014.

\bibitem{abrikosov1963methods}
AA~Abrikosov, LP~Gorkov, and IE~Dzyaloshinski.
\newblock {\em Methods of Quantum Field Theory in Statistical Physics}.
\newblock Dover Publications, 1963.

\bibitem{zinn1996quantum}
Jean Zinn-Justin.
\newblock {\em Quantum field theory and critical phenomena}.
\newblock Clarendon Press, 1996.

\bibitem{kleinert2001critical}
Hagen Kleinert and Verena Schulte-Frohlinde.
\newblock {\em Critical Properties of $\phi^4$-theories}.
\newblock World Scientific, 2001.

\bibitem{kleinert2009path}
Hagen Kleinert.
\newblock {\em Path integrals in quantum mechanics, statistics, polymer
  physics, and financial markets}.
\newblock World scientific, 2009.

\bibitem{metzner1991linked}
Walter Metzner.
\newblock Linked-cluster expansion around the atomic limit of the hubbard
  model.
\newblock {\em Physical Review B}, 43(10):8549, 1991.

\bibitem{irving1984methods}
AC~Irving and CJ~Hamer.
\newblock Methods in hamiltonian lattice field theory (ii). linked-cluster
  expansions.
\newblock {\em Nuclear Physics B}, 230(3):361--384, 1984.

\bibitem{gelfand1990perturbation}
Martin~P Gelfand, Rajiv~RP Singh, and David~A Huse.
\newblock Perturbation expansions for quantum many-body systems.
\newblock {\em Journal of Statistical Physics}, 59(5-6):1093--1142, 1990.

\bibitem{fetter2012quantum}
Alexander~L Fetter and John~Dirk Walecka.
\newblock {\em Quantum theory of many-particle systems}.
\newblock Courier Corporation, 2012.

\bibitem{mahan2013many}
Gerald~D Mahan.
\newblock {\em Many-particle physics}.
\newblock Springer Science \& Business Media, 2013.

\bibitem{bender1999advanced}
Carl~M Bender and Steven~A Orszag.
\newblock {\em Advanced mathematical methods for scientists and engineers I:
  Asymptotic methods and perturbation theory}.
\newblock Springer Science \& Business Media, 1999.

\bibitem{pelster2003high}
Axel Pelster, Hagen Kleinert, and Michael Schanz.
\newblock High-order variational calculation for the frequency of time-periodic
  solutions.
\newblock {\em Physical Review E}, 67(1):016604, 2003.

\bibitem{vidanovic2011nonlinear}
Ivana Vidanovi{\'c}, Antun Bala{\v{z}}, Hamid Al-Jibbouri, and Axel Pelster.
\newblock Nonlinear bose-einstein-condensate dynamics induced by a harmonic
  modulation of the s-wave scattering length.
\newblock {\em Physical Review A}, 84(1):013618, 2011.

\bibitem{al2013geometric}
Hamid Al-Jibbouri, Ivana Vidanovi{\'c}, Antun Bala{\v{z}}, and Axel Pelster.
\newblock Geometric resonances in bose--einstein condensates with two-and
  three-body interactions.
\newblock {\em Journal of Physics B: Atomic, Molecular and Optical Physics},
  46(6):065303, 2013.

\bibitem{pollet2009absence}
L~Pollet, NV~Prokof’ev, BV~Svistunov, and M~Troyer.
\newblock Absence of a direct superfluid to mott insulator transition in
  disordered bose systems.
\newblock {\em Physical review letters}, 103(14):140402, 2009.

\bibitem{graham2009order}
Robert Graham and Axel Pelster.
\newblock Order via nonlinearity in randomly confined bose gases.
\newblock {\em International Journal of Bifurcation and Chaos},
  19(08):2745--2753, 2009.

\bibitem{khellil2016hartree}
Tama Khellil and Axel Pelster.
\newblock Hartree--fock mean-field theory for trapped dirty bosons.
\newblock {\em Journal of Statistical Mechanics: Theory and Experiment},
  2016(6):063301, 2016.

\bibitem{khellil2016analytical}
Tama Khellil, Antun Bala{\v{z}}, and Axel Pelster.
\newblock Analytical and numerical study of dirty bosons in a
  quasi-one-dimensional harmonic trap.
\newblock {\em New Journal of Physics}, 18(6):063003, 2016.

\bibitem{khellil2017dirty}
Tama Khellil and Axel Pelster.
\newblock Dirty bosons in a three-dimensional harmonic trap.
\newblock {\em Journal of Statistical Mechanics: Theory and Experiment},
  2017(9):093108, 2017.

\end{thebibliography}

\end{document}